\documentclass[11pt]{article}

\usepackage[a4paper]{geometry}

\usepackage[latin1]{inputenc}

\usepackage{latexsym}
\usepackage{amsmath}
\usepackage{amsthm}
\usepackage{amssymb}

\newtheorem{theorem}{Theorem}
\newtheorem{lemma}{Lemma}
\newtheorem{corollary}{Corollary}
\newtheorem{proposition}{Proposition}
\theoremstyle{remark}
\newtheorem*{remark}{Remark}

\title{Static cylindrically symmetric spacetimes}

\author{Mikael Fjällborg}

\begin{document}

\maketitle
\textbf{Abstract}
We prove the existence of static solutions to the cylindrically symmetric Einstein-Vlasov system, and we show that the matter cylinder has finite extension in two of the three spatial dimensions. The same results are also proved for a quite general class of equations of state for perfect fluids coupled to the Einstein equations, extending the class of equations of state considered in \cite{BL}. We also obtain this result for the Vlasov-Poisson system.

\bigskip

\section{Introduction}
The most frequently studied static spacetimes containing matter are the spherically symmetric ones. This class of spacetimes is compatible with asymptotic flatness. Another interesting class of asymptotically flat spacetimes are the axially symmetric ones. However, a study of this class is at present out of reach. As a first possible step towards axial symmetry it could be useful to study the cylindrically symmetric spacetimes, although cylindrical symmetry is not compatible with asymptotic flatness, nor are there any cylindrical configurations in the universe. Nevertheless, cylindrically symmetric spacetimes can be considered as a subclass of the axially symmetric ones with an extra symmetry requirement. It should also be pointed out that time dependent cylindrically symmetric spacetimes admit gravitational waves, in contrast to the spherically symmetric case, and still reduce the complexity of the Einstein equations. 

In 1917 Levi-Civita obtained the most general static cylindrically symmetric vacuum solutions. Since then many explicit cylindrical solutions have been obtained in the fluid case, cf. \cite{KSHM}. Unfortunately these solutions are often local, and no global analysis is usually available. However, in \cite{BL} the global properties of static cylindrically symmetric spacetimes with perfect fluid matter is studied, and global existence of solutions and finiteness of the radius of the fluid cylinder is shown. It is shown that when the fluid cylinder has finite extension, it is possible to glue it smoothly with a Levi-Civita solution, and in this way obtain a global solution. To prove the finite extension of the fluid cylinder the authors make use of the fact that the boundary density, i.e. the energy density of the fluid on the boundary where the pressure vanishes, is positive in an essential way. The main part of this paper is concerned with the Einstein-Vlasov system. However, we will also improve previous results for perfect fluids by including all equations of state that satisfy

\begin{equation}
\int_{0}^{P_{c}}\frac{dp}{\rho (p) +p}<\infty,\label{eta1}
\end{equation}

where $\rho $ is the energy density, $p$ the pressure and $P_{c}$ the central pressure. The Einstein-Vlasov system is the Einstein equations coupled to a collisionless kinetic matter model, cf. \cite{A,R} for an introduction to the Einstein-Vlasov system. Prior to this study, only the spherically symmetric Einstein-Vlasov system has been studied in the static case, cf. \cite{FHU,Rn,RR1,RR}. However, in the time dependent situation more symmetry classes have been studied, cf. \cite{A} for a review of the results on the Einstein-Vlasov system. Cylindrically symmetric time dependent solutions have been considered in \cite{F}.
 
For the static spherically symmetric Einstein-Vlasov system the proof of local existence is straightforward, but the proof of global existence is more involved. The proof of finiteness of extension of matter is essentially done in one of the following ways. Either it is proved  by using the non-relativistic limit, knowing that this limit has finite extension of matter, or it is proved by introducing new variables in an appropriate way, and by showing that these functions blow up at a finite radius one concludes that the matter has finite extension, cf. \cite{Rn} and \cite{RR} for the different approaches. We will rely on the last approach. We point out that in \cite{FHU} a completely different approach is used to investigate the static spherically symmetric Einstein-Vlasov system. In \cite{FHU} the system of equations is transformed to an autonomous dynamical system on a state space with compact closure, and the solution orbits are studied to extract information whether the corresponding solutions have finite radius or not.

For the static cylindrically symmetric Einstein-Vlasov system the proof of local existence is straightforward, but uses the fact that the matter terms are continuously differentiable. To ensure this differentiability property and to be able to compare our results with the spherically symmetric case, and also to be able to handle the matter terms more easily, we study the following class of distribution functions. Let $f(r,v):=\phi (E)L^{l}$, $l>-1$, where $E$ is the particle energy and $L$ is the modulus of the angular momentum, cf. section 2. We also require that $\phi\in C^{1}((0,E_{0} ))$, where $E_{0}>0$ is a constant, a cut-off energy. To obtain global existence we need to do a careful analysis of the equations involved. The proof of finite extension of matter do not need the asymptotic structure of the distribution function, in contrast to the spherically symmetric case, where it is needed. Heuristically, one of the reasons that the proof of finiteness of extension is easier in the cylindrically symmetric case than in the spherically symmetric, is that the singularity of the equations at $r=0$ is stronger in the latter. 

As mentioned above we also prove the same result for the static cylindrically symmetric Einstein equations coupled to a perfect fluid which satisfies equation $\left (\ref{eta1}\right )$. In the analogous results for the spherically symmetric case the asymptotically polytropic behavior as the pressure tends to zero is taken into account, which is not needed for the cylindrically symmetric case. This is in turn analogous to the comparison between the cylindrically- and spherically symmetric Einstein-Vlasov system as mentioned above. We also show that the matter solutions, both for Vlasov matter and perfect fluids, can be glued with a unique vacuum solution. 
 
In the final part of the paper we also consider the non-relativistic Vlasov-Poisson system. Global existence have been proved previously in \cite{BB} under more general conditions than we have, but we show in addition that the matter cylinder has finite extension. 

\section{Static cylindrical spacetimes}

Consider a spacetime with two hypersurface orthogonal Killing vectors, where one is a translation $\xi_{z}$, and the other a rotation $\xi_{\theta}$. Let the spacetime also be equipped with a timelike Killing vector $\xi_{t}$. Then the most general metric can be written in the form, cf. \cite{KSHM},

\begin{equation*}
ds^{2}=-e^{2\nu}dt^{2}+e^{2(\gamma -\psi )}dr^{2}+e^{2\psi}dz^{2}+r^{2}e^{-2\psi}d\theta ^{2},
\end{equation*}

where $t\in (-\infty,\infty )$, $r\in [0,\infty )$, $z\in (-\infty,\infty )$ and $\theta\in [0,2\pi )$, and where $\nu$, $\gamma$ and $\psi$ only depend on $r$. We normalize the gravitational constant $G=1$, and the speed of light $c=1$. Then the Einstein matter equations read, cf. \cite{SST},

\begin{equation}
\frac{\gamma '}{r}=\psi '^{2}-8\pi e^{2(\gamma-\psi)}T_{t}^{t}, \label{E1}
\end{equation}

\begin{equation}
\frac{\nu '+\psi '}{r}=\psi '^{2}+8\pi e^{2(\gamma-\psi)}T_{r}^{r}, \label{E2}
\end{equation}

\begin{equation}
\psi ''+\frac{\psi '}{r}+\psi '8\pi re^{2(\gamma-\psi )}(T_{r}^{r}+T_{t}^{t})=4\pi e^{2(\gamma-\psi)}(T_{\theta}^{\theta}-T_{z}^{z}+T_{r}^{r}+T_{t}^{t}), \label{E3}
\end{equation}

\begin{equation}
\nu ''+\frac{\nu '}{r}+\nu '8\pi re^{2(\gamma-\psi)}(T_{r}^{r}+T_{t}^{t})=4\pi e^{2(\gamma -\psi )}(T_{r}^{r}+T_{\theta }^{\theta }+T_{z}^{z}-T_{t}^{t}), \label{E4}
\end{equation}

where $T_{i}^{i},\text{ } i=t,r,\theta$ and $z$ are the components of the energy-momentum tensor. For a perfect fluid the nonzero components of the energy-momentum tensor read
\begin{equation*}
T_{t}^{t}=-\rho ,\text{ }T_{r}^{r}=T_{z}^{z}=T_{\theta}^{\theta}=P,
\end{equation*}
where $\rho$ is the energy density and $P$ is the pressure. From $\nabla_{a}T_{b}^{a}=0$ we obtain

\begin{equation}
\frac{dP}{dr}=-\nu '(\rho +P). \label{E5}
\end{equation}

In kinetic theory the energy-momentum tensor is defined by 

\begin{equation*}
T_{ab}(x)=-\int_{\mathbb{R}^{3}}f(x,p)\frac{p_{a}p_{b}}{p_{0}}\sqrt{|g|}dp,
\end{equation*}

where $a,b=0,1,2,3$ and $|g|$ is the determinant of the spacetime metric, and $x$ are the spatial coordinates with corresponding momentum $p$, i.e. $p$ is the spatial part of the four-momentum. Furthermore $p^{0}$ is expressed through $p^{a}$ and $x$ via $g_{ab}p^{a}p^{b}=-1$, where we have assumed that all particles have the same rest mass $m=1$. We introduce new momentum variables $v^{a},\text{ } a=0,1,2,3$ by

\begin{equation*}
v^{0}=e^{\nu}p^{0},\text{ } v^{1}=e^{\gamma -\psi}p^{1},\text{ } v^{2}=e^{\psi}p^{2},\text{ } v^{3}=re^{-\psi}p^{3},
\end{equation*}
and note that
\begin{equation*}
v^{0}=\sqrt{1+(v^{1})^{2}+(v^{2})^{2}+(v^{3})^{2}}.
\end{equation*}
 
The nonzero components of the energy-momentum tensor for static cylindrically symmetric kinetic matter then read

\begin{equation*}
T_{t}^{t}=-\int_{\mathbb{R}^{3}}f(r,v)v^{0}dv:=-\rho,
\end{equation*}

\begin{equation*}
T_{i}^{i}=\int_{\mathbb{R}^{3}}f(r,v)\frac{(v^{i})^{2}}{v^{0}}dv:=P_{i},\text{ } i=1,2,3.
\end{equation*}

We immediately obtain the following important inequality
\begin{equation}
\rho >P_{1}+P_{2}+P_{3}, \label{energi}
\end{equation}
which holds on the $r$ support of $f$.
The distribution function $f$ satisfies the Vlasov equation

\begin{equation*}
\frac{v^{1}}{v^{0}}\frac{\partial f}{\partial r}-[\nu 'v^{0}-\psi '\frac{(v^{2})^{2}}{v^{0}}+(\psi '-\frac{1}{r})\frac{(v^{3})^{2}}{v^{0}}]\frac{\partial f}{\partial v^{1}}
\end{equation*}
\begin{equation}
-\psi '\frac{v^{1}v^{2}}{v^{0}}\frac{\partial f}{\partial v^{2}}-(\frac{1}{r}-\psi ')\frac{v^{1}v^{3}}{v^{0}}\frac{\partial f}{\partial v^{3}}=0. \label{vl}
\end{equation}

Now multiply the Vlasov equation $\left (\ref{vl}\right )$ by $v^{1}$ and integrate in velocity space, assuming compact support in the velocity variables, or at least sufficiently fast decay to zero as $|v|\rightarrow\infty $, then we obtain

\begin{equation}
\frac{dP_{1} }{dr} = -\nu '(\rho +P_{1})-l\psi 'P_{1}+\frac{l}{r}P_{1}. \label{pe}
\end{equation}

Note that this reduces to equation $\left (\ref{E5}\right )$ in the particular case of a locally isotropic distribution function, i.e. the distribution function only depends on the energy, see below, since then $P_{1}=P_{2}=P_{3}=:P$.

We have three constants of motion. The particle energy $E=e^{\nu}v^{0}$, the angular momentum squared $L^{2}=r^{2}e^{-2\psi}|v^{3}|^{2}$ and $Z=e^{\psi}v^{2}$. If $f(r,v)=\Phi (E,L,Z)$, with $\Phi\in C^{1}$, then $f$ satisfies the Vlasov equation. We will however restrict the function class and only consider functions of the type

\begin{equation*}
f(r,v)=\phi(E)L^{l},\text{ } l>-1.
\end{equation*}
Furthermore, $\phi \in C^{1}((0,E_{0})) $, $E_{0}>0$ and $k>-1$ such that on every compact subset $K\subset (0,\infty )$ there exists a constant $C\geq 0$ such that $0\leq\phi (E)\leq C(E_{0}-E)_{+}^{k}$. The matter terms will be continuously differentiable under the assumption $k+\frac{l+1}{2}>0$, cf. Lemma 1. Then the matter terms read

\begin{equation}
\rho (r)=\frac{4\pi r^{l}e^{-(l+4)\nu-l\psi}}{l+1}\int_{e^{\nu}}^{E_{0}}\phi (E)E^{2}(E^{2}-e^{2\nu})^{\frac{l+1}{2}}dE, \label{rho1}
\end{equation}

\begin{equation}
P_{i}(r)=4\pi A_{i}r^{l}e^{-(l+4)\nu-l\psi}\int_{e^{\nu}}^{E_{0}}\phi (E)(E^{2}-e^{2\nu})^{\frac{l+3}{2}}dE,\text{ } i=1,2,3, \label{pi}
\end{equation}

where 

\begin{equation*}
A_{1}=A_{2}=\frac{1}{(l+1)(l+3)},
\end{equation*}

and

\begin{equation*}
A_{3}=\frac{1}{l+3}.
\end{equation*}

Now define

\begin{equation*}
H_{m}(u):=\int_{u}^{E_{0}}\phi (E)(E^{2}-u^{2})^{m}dE,
\end{equation*}

then $\left (\ref{rho1}\right )$ and $\left (\ref{pi}\right )$ become

\begin{equation}
\rho (r)=\frac{4\pi r^{l}e^{-(l+4)\nu -l\psi}}{l+1}[H_{\frac{l+3}{2}}(e^{\nu})+e^{2\nu}H_{\frac{l+1}{2}}(e^{\nu})],\label{rho}
\end{equation}

\begin{equation}
P_{i}(r)=4\pi A_{i}r^{l}e^{-(l+4)\nu-l\psi}H_{\frac{l+3}{2}}(e^{\nu}),\text{ } i=1,2,3.\label{P}
\end{equation}

The boundary conditions which supplement equations $\left (\ref{E1}\right )$-$\left (\ref{E4}\right )$ are given by

\begin{equation}
\gamma =\frac{d\gamma}{dr}=\nu =\frac{d\nu}{dr}=\psi =\frac{d\psi}{dr}=0,\text{ at }r=0, \label{BC}
\end{equation}

and follow from the regularity on the axis $r=0$, cf. \cite{BL}. 

We will also consider the non-relativistic counterpart to the Einstein-Vlasov system, namely the Vlasov-Poisson system which in the static cylindrically symmetric case reads

\begin{equation}
\frac{1}{r}(rU'(r))'=4\pi\rho, \label{vp1}
\end{equation}

\begin{equation}
v\cdot\frac{\partial f}{\partial x}-\nabla U\cdot\frac{\partial f}{\partial v}=0. \label{vp2}
\end{equation}

Here $\rho$ is given by $\rho (r):=\int_{\mathbb{R}^{3}}f(r,v)dv$. The conserved quantities in the non-relativistic setting are the energy $E:=\frac{1}{2}|v|^{2}+U(r)$, the angular momentum squared $L^{2}:=r^{2}|u|^{2}\sin ^{2}\theta $, where $u:=(v^{1},v^{2})$ and $\theta $ the angle between $x$ and $u$ in the plane of fixed $z$, and the quantity $Z:=v^{3}$. As in the relativistic case we have that any reasonable function which only depends on these variables satisfies the Vlasov equation. By restricting to the same class of functions as before, i.e. $f(r,v):=\phi (E)L^{l}$, $l>-1$, $\phi\in C^{1}((-\infty ,E_{0}))$, $E_{0}>-\infty $ a constant, and $\phi $ satisfying the requirements in Lemma 1 below, we obtain

\begin{equation}
\rho (r)=\frac{2^{\frac{l+5}{2}}\pi}{l+1}r^{l}g_{\frac{l+1}{2}}(U(r)), \label{vp3}
\end{equation}
where 

\begin{equation}
g_{m}(u):=\int_{u}^{E_{0}}\phi (E)(E-u)^{m}dE.\label{vp4}
\end{equation}

Define $E_{min}:=-\infty$ in the non-relativistic case and $E_{min}:=0$ in the relativistic case. The following lemma will be needed later, for a proof cf. \cite{RR}.

\begin{lemma} \cite{RR} Let $\phi :(E_{min},\infty )\rightarrow\mathbb{R}$ be measurable, $E_{0}>E_{min}$, and $k>-1$ such that on every compact subset $K\subset (E_{min},\infty )$ there exists $C\geq 0$ such that $0\leq\phi (E)\leq C(E_{0}-E)^{k}_{+}, E\in K$. Let $H_{m}(u)$ and $g_{m}(u)$ be defined as above i.e.

\begin{equation*}
H_{m}(u)=\int_{u}^{E_{0}}\phi (E)(E^{2}-u^{2})^{m}dE, u\in (0,\infty ),\label{me}
\end{equation*}

\begin{equation*}
g_{m}(u):=\int_{u}^{E_{0}}\phi (E)(E-u)^{m}dE, u\in(-\infty ,\infty ).
\end{equation*}

If $m>-1$ and $k+m+1>0$ then $g_{m},H_{m}\in C((E_{min},\infty ))$.
If $m>0$ and $k+m>0$ then $g_{m},H_{m}\in C^{1}((E_{min},\infty ))$ with 

\begin{equation*}
g_{m}'=-mg_{m-1},
\end{equation*}

\begin{equation*}
H_{m}'=-2muH_{m-1}.
\end{equation*}
\end{lemma}

With a distribution function $\phi $ as in Lemma 1, we have that when $u\geq E_{0}$, then $H_{m}(u)=g_{m}(u)=0$, which imply that the matter terms vanish.

\section{Existence and finite extension of matter}
\subsection{The Einstein-Vlasov system}

For later purposes it is convenient to rewrite the Einstein equations with Vlasov matter as follows. We follow \cite{SST} and introduce the new variable 
\begin{equation}
M:=\frac{1}{8}(1-e^{2(\nu +\psi -\gamma)})\label{M}.
\end{equation} 
The reason for this transformation is that the system then has a structure similar to the static spherically symmetric Einstein-Vlasov system, with $M(r)$ used in a similar way as the local ADM mass $m(r)$ in the spherically symmetric case, cf. \cite{RR}. Note that cylindrical spacetimes cannot be asymptotically flat and hence do not have any ADM mass.
The Einstein matter equations are transformed to

\begin{equation}
\frac{dM}{dr}=2\pi re^{2\nu}(\rho -P_{1}), \label{s1}
\end{equation}

\begin{equation}
\frac{d\psi}{dr}=\frac{4\pi}{r\sqrt{1-8M}}\int_{0}^{r}\tilde{r}e^{2\nu}\frac{1}{\sqrt{1-8M}}(P_{3}-\rho )d\tilde{r}, \label{s2}
\end{equation}

\begin{equation}
 \frac{d\nu}{dr}=\frac{4\pi}{r\sqrt{1-8M}}\int_{0}^{r}\tilde{r}e^{2\nu}\frac{1}{\sqrt{1-8M}}(P_{3}+2P_{1}+\rho )d\tilde{r}, \label{s3}
\end{equation}
where we have used the regularity at the axis.

The boundary conditions follow from equation $\left (\ref{BC}\right )$ and read

\begin{equation}
M=\frac{dM}{dr}=\nu=\frac{d\nu}{dr}=\psi=\frac{d\psi}{dr}=0,\text{ at } r=0. \label{B}
\end{equation}

We have the following local (in $r$) existence theorem.

\begin{theorem} 
Let $f(r,v)=\phi (E)L^{l}, l>-1$ and $\phi$ as in Lemma 1 with $k+\frac{l+1}{2}>0$ so that $H_{m}\in C^{1}((0,\infty ))$, $m=\frac{l+1}{2},\frac{l+3}{2}$. Then for $\delta >0$ sufficiently small, there exists a unique $C^{2}$-solution, $(\gamma ,\psi ,\nu )$, to the system $\left (\ref{E1}\right )$-$\left (\ref{E4}\right )$, $\left (\ref{vl}\right )$ with boundary conditions $\left (\ref{BC}\right )$, on the interval $[0,\delta]$.
\end{theorem}

Theorem 1 follows immediately from the next lemma, since the system $\left (\ref{E1}\right )$-$\left (\ref{E4}\right )$, $\left (\ref{vl}\right )$ with boundary conditions $\left (\ref{BC}\right )$ is equivalent to system $\left (\ref{s1}\right )$-$\left( \ref{s3}\right)$, $\left (\ref{vl}\right )$ with boundary conditions $\left (\ref{B}\right )$.

\begin{lemma} 
Let $f$ and $\phi$ be as above, then for $\delta >0$ sufficiently small, there exists a unique $C^{1}$-solution $(M,\psi ,\nu )$ to the system $\left (\ref{s1}\right )-\left (\ref{B}\right )$ on the interval $[0,\delta]$.
\end{lemma}

\begin{proof}
Take $\delta >0$. Define the set

\begin{equation*}
B=\{w=(M,\psi ,\nu )\in C^{0}([0,\delta],\mathbb{R}^{3}):M(0)=0,\psi (0)=0,\nu (0)=0,
\end{equation*}
\begin{equation*}
\parallel M\parallel_{\infty}\leq K_{1},\parallel \psi \parallel_{\infty}\leq K_{2}, \parallel \nu \parallel_{\infty}\leq K_{3}\},
\end{equation*}  
 
where $K_{i},i=1,2,3$ are positive constants, $K_{1}<\frac{1}{8}$, and $B$ is equipped with the norm $\parallel w\parallel _{B}:=\parallel M\parallel _{\infty }+\parallel\psi \parallel _{\infty}+\parallel\nu\parallel _{\infty }$. $B$ is a closed set in the Banach space $C^{0}([0,\delta ],\mathbb{R}^{3})$, and thus a complete metric space.

Now define the operator $T$, acting on $B$ by

\begin{equation*}
Tw:=\Bigl ( (Tw)_{1},(Tw)_{2},(Tw)_{3}\Bigr ),
\end{equation*}

where

\begin{equation}
(Tw)_{1}:=2\pi\int_{0}^{r} \tilde{r}e^{2\nu}(\rho -P_{1})d\tilde{r},\label{Tr1}
\end{equation}
\begin{equation}
(Tw)_{2} :=\int_{0}^{r}\frac{4\pi}{\tilde{r}\sqrt{1-8M}}\int_{0}^{\tilde{r}}\hat{r}e^{2\nu }\frac{1}{\sqrt{1-8M}}(P_{3}-\rho)d\hat{r}d\tilde{r},\label{Tr2}
\end{equation}
\begin{equation}
(Tw)_{3} :=\int_{0}^{r}\frac{4\pi}{\tilde{r}\sqrt{1-8M}}\int_{0}^{\tilde{r}}\hat{r}e^{2\nu }\frac{1}{\sqrt{1-8M}}(P_{3}+2P_{1}+\rho)d\hat{r}d\tilde{r}.\label{Tr3}
\end{equation}

Recall that the matter terms $\rho , P_{1},P_{2},P_{3}$ are functionals of the metric components and hence has a $w$ dependence. First we show that $T(B)\subset B$. Take $w\in B$. Since $w\in B$, we have that $e^{\nu}\in [e^{-K_{3}},e^{K_{3}}]$. Recall by Lemma 1, $H_{\frac{l+1}{2}},H_{\frac{l+3}{2}}\in C^{1}((0,\infty ))$, and since the arguments take values in a closed bounded interval which do not contain zero, $H_{\frac{l+1}{2}}$ ,$H_{\frac{l+3}{2}}$, $H'_{\frac{l+1}{2}}$ and $H'_{\frac{l+3}{2}}$ are all bounded. The argument is similar for all the components of the mapping, so we only show that the second component of the mapping satisfies the claim above. Since $l>-1$, equations $\left (\ref{rho}\right )$ and $\left (\ref{P}\right )$ yields the estimate

\begin{equation*}
\big | \hat{r}e^{2\nu }\frac{1}{\sqrt{1-8M}}(P_{3}-\rho )\big |\leq C\hat{r}^{l+1}.
\end{equation*}
This implies that
\begin{equation*}
\Big |\frac{4\pi}{\tilde{r}\sqrt{1-8M}}\int_{0}^{\tilde{r}}\hat{r}e^{2\nu }\frac{1}{\sqrt{1-8M}}(P_{3}-\rho)d\hat{r}\Big |\leq C\tilde{r}^{l+1},
\end{equation*}
which yields
\begin{equation*}
\parallel (Tw)_{2}\parallel _{\infty }\leq Cr^{l+2}.
\end{equation*}
Similar estimates  yield that $\parallel (Tw)_{i}\parallel _{B}\leq Cr^{l+2}$, $i=1,2,3$. Hence, for $\delta $ small enough $Tw\in B$, and since $w$ was arbitrary $T(B)\subset B$. We point out that the constant $C$ above may change from line to line and does not depend on $\delta $.

We now claim that $T$ acts as a contraction on the space $B$ for $\delta$ sufficiently small. Then, by the Banach contraction theorem we have a unique continuous solution to the system $\left (\ref{s1}\right )$-$\left (\ref{B}\right )$, but it is then straightforward to verify that this solution in fact is a $C^{1}$-solution. We want to estimate $\parallel Tw-T\hat{w}\parallel _{B}$ where $w,\hat{w}\in B$. Since the estimates of the different terms are similar we just sketch the proof of the second term. The mean value theorem will be used to estimate the differences, and Lemma 1 will be used to ensure the continuity of the matter terms and its derivatives. Now the difference in the second component of the mapping $T$ can be written

\begin{equation*}
|(Tw)_{2} -(T\hat{w})_{2}|=\bigg |\int_{0}^{r}\frac{4\pi}{\tilde{r}\sqrt{1-8M}}\int_{0}^{\tilde{r}}\hat{r}e^{2\nu}\frac{1}{\sqrt{1-8M}}(P_{3}-\rho )d\hat{r}d\tilde{r}
\end{equation*}
\begin{equation*}
-\int_{0}^{r}\frac{4\pi}{\tilde{r}\sqrt{1-8\hat{M}}}\int_{0}^{\tilde{r}}\hat{r}e^{2\hat{\nu}}\frac{1}{\sqrt{1-8\hat{M}}}(\hat{P}_{3}-\hat{\rho })d\hat{r}d\tilde{r}\bigg |
\end{equation*}

\begin{equation*}
=\bigg |\int_{0}^{r}\bigg [\frac{4\pi }{\tilde{r}}(\frac{1}{\sqrt{1-8M}}-\frac{1}{\sqrt{1-8\hat{M}}})\int_{0}^{\tilde{r}}\hat{r}e^{2\nu}\frac{1}{\sqrt{1-8M}}(P_{3}-\rho )d\hat{r}
\end{equation*}
\begin{equation*}
+\frac{4\pi }{\tilde{r}\sqrt{1-8\hat{M}}} [\int_{0}^{\tilde{r}} \hat{r}[(e^{2\nu}-e^{2\hat{\nu}})\frac{1}{\sqrt{1-8M}}+e^{2\hat{\nu}}(\frac{1}{\sqrt{1-8M}}-\frac{1}{\sqrt{1-8\hat{M}}})](P_{3}-\rho )d\hat{r}
\end{equation*}

\begin{equation*}
+\int_{0}^{\tilde{r}} \hat{r}e^{2\hat{\nu}}\frac{1}{\sqrt{1-8\hat{M}}}(P_{3}-\hat{P}_{3}-\rho +\hat{\rho })d\hat{r}]\bigg ]d\tilde{r}\bigg |,
\end{equation*}

where $\hat{P}_{3}$ and $\hat{\rho }$ indicates the dependence on $\hat{w}$.
 Define
\begin{equation*}
F(\nu,\psi):=e^{-[(l+2){\nu } +l{\psi } ]},
\end{equation*}
then, since $w,\hat{w}\in B$, by the mean value theorem

\begin{equation*}
|F(\nu ,\psi )-F(\hat{\nu },\hat{\psi })|\leq\underset{\tilde{\nu },\tilde{\psi }\in B}{\sup }\displaystyle\left(e^{-[(l+2)\tilde{\nu }+l\tilde{\psi }]}\sqrt{(l+2)^{2}+l^{2}}\displaystyle\right)|(\nu ,\psi )-(\hat{\nu },\hat{\psi })|
\end{equation*}

\begin{equation*}
\leq C\parallel w-\hat{w}\parallel _{B}.
\end{equation*}

Now define 
\begin{equation*}
G(M):=\frac{1}{\sqrt{1-8M}},
\end{equation*}
and again by the mean value theorem
\begin{equation*}
\bigl |G(M)-G(\hat{M})\bigr |\leq C\parallel M-\hat{M}\parallel _{\infty }
\end{equation*}
\begin{equation*}
\leq C\parallel w-\hat{w}\parallel _{B}.
\end{equation*}

By using equations $\left (\ref{rho}\right )$ and $\left (\ref{P}\right )$ we obtain as before that
\begin{equation*}
\bigl |\tilde{r}e^{2\nu }\frac{1}{\sqrt{1-8M}}(P_{3} -\rho )\bigr |\leq C\tilde{r}^{l+1}.
\end{equation*}
Hence, 
\begin{equation*}
\bigg |\int_{0}^{r}\frac{4\pi }{\tilde{r}}(\frac{1}{\sqrt{1-8M}}-\frac{1}{\sqrt{1-8\hat{M}}})\int_{0}^{\tilde{r}}\hat{r}e^{2\nu }\frac{1}{\sqrt{1-8M}}(P_{3}-\rho )d\hat{r}d\tilde{r}\bigg |
\end{equation*}
\begin{equation}
\leq Cr^{l+2}\parallel w-\hat{w}\parallel _{B}. \label{T2a}
\end{equation}
 Similarly by using the mean value theorem we have 
\begin{equation*}
\big |e^{2\nu}-e^{2\hat{\nu }}\big |\leq C\parallel w-\hat{w}\parallel _{B},
\end{equation*}
which yields
\begin{equation*}
\bigg | \int_{0}^{r}\frac{4\pi }{\tilde{r}\sqrt{1-8\hat{M}}}\int_{0}^{\tilde{r}}\hat{r}\frac{1}{\sqrt{1-8\hat{M}}}(P_{3}-\rho )(e^{2\nu }-e^{2\hat{\nu }})d\hat{r}d\tilde{r}\bigg |
\end{equation*}
\begin{equation}
\leq Cr^{l+2}\parallel w-\hat{w}\parallel _{B}, \label{T2b}
\end{equation}

and 
\begin{equation*}
\bigg |\int_{0}^{r}\frac{4\pi }{\tilde{r}\sqrt{1-8\hat{M}}}\int_{0}^{\tilde{r}}\hat{r}e^{2\hat{\nu }}(P_{3}-\rho )(\frac{1}{\sqrt{1-8M}}-\frac{1}{\sqrt{1-8\hat{M}}})d\hat{r}d\tilde{r}\bigg |
\end{equation*}
\begin{equation}
\leq Cr^{l+2}\parallel w-\hat{w}\parallel _{B}. \label{T2c}
\end{equation}

By the mean value theorem again
\begin{equation*}
\big |H_{m}(e^{\nu })-H_{m}(e^{\hat{\nu }})\big |\leq\underset{\tilde{\nu }\in B}{\sup }|H'_{m}(e^{\tilde{\nu }})||e^{\nu}-e^{\hat{\nu }}|
\end{equation*}
\begin{equation*}
\leq C\parallel w-\hat{w}\parallel _{B},
\end{equation*}

with $m=\frac{l+1}{2}$ or $m=\frac{l+3}{2}$, which implies that

\begin{equation*}
\bigg |\int_{0}^{r}\frac{4\pi }{\tilde{r}\sqrt{1-8\hat{M}}}\int_{0}^{\tilde{r}}\hat{r}e^{2\hat{\nu }}\frac{1}{\sqrt{1-8\hat{M}}}(P_{3}-\hat{P}_{3}-\rho +\hat{\rho })d\hat{r}d\tilde{r}\bigg |
\end{equation*}

\begin{equation}
\leq Cr^{l+2}\parallel w-\hat{w}\parallel _{B}. \label{T2d}
\end{equation}

Equations $\left (\ref{T2a}\right )-\left (\ref{T2d}\right )$ yield

\begin{equation*}
\parallel (Tw)_{2} -(T\hat{w})_{2}\parallel _{\infty }\leq Cr^{l+2}\parallel w-\hat{w}\parallel _{B}.
\end{equation*}

Similar estimates hold for the first and third component of the mapping $T$. Hence for $\delta >0$ small enough, $T$ is a contraction mapping on $B$ since $l+2>1$. Thus, there exists a unique continuous solution to the system $\left (\ref{s1}\right )-\left (\ref{s3}\right )$ on the interval $[0,\delta ]$, with $\psi (0)=\nu (0)=M(0)=0$. Since $l>-1$ we obtain by using the equations $\left (\ref{s1}\right )$-$\left (\ref{s3}\right )$ that the solution in fact is a $C^{1}$-solution that verifies the boundary conditions $\left (\ref{B}\right )$.  

\end{proof}

\begin{remark} By using equations $\left (\ref{E1}\right )$-$\left (\ref{E4}\right )$ we immediately obtain that $(\nu ,\psi ,\gamma )$ is a $C^{2}$-solution.
\end{remark}

Define 
\begin{equation*}
\tilde{P}:=4\pi r^{l}e^{-(l+4)\nu -l\psi}H_{\frac{l+3}{2}}(e^{\nu }),
\end{equation*}
then $P_{i}=A_{i}\tilde{P}$, $i=1,2,3$. The following lemma gives us some useful information about the behavior of the matter and the metric. 

\begin{lemma} 
Suppose we have a $C^{1}$-solution $(M,\psi ,\nu )$ to the system $\left (\ref{s1}\right )$-$\left (\ref{B}\right )$ with $f(r,v)=\phi (E)L^{l},\text{ } l>-1,\text{ } k+\frac{l+1}{2}>0$ and $\phi $ as in Lemma 1, then $M$, $\nu$ and $\gamma$ are strictly increasing and $\psi$ is strictly decreasing as long as $\rho >0$ and $M<\frac{1}{8}$.
\end{lemma}
\begin{proof}
From equations $\left (\ref{s1}\right )$, $\left (\ref{E1}\right )$ and $\left (\ref{s3}\right )$ we immediately obtain that $M$, $\gamma$ and $\nu $ are strictly increasing since $\rho >0$. Furthermore from equation $\left (\ref{s2}\right )$ we obtain that $\psi $ is strictly decreasing. 
\end{proof}

Define 
\begin{equation*}
R_{ex}:=\text{maximal interval of existence},
\end{equation*}
and

\begin{equation*}
R:=\sup\{r\leq R_{ex}:\rho (\tilde{r})>0,\text{ for }\tilde{r}\in [0,r)\}.
\end{equation*}

\begin{lemma} Assume that there exists $r_{0}\in (0,R_{ex})$ such that $\rho (r_{0})=0$, then $\rho (r)=0$ for $r\in [r_{0},R_{ex})$.
\end{lemma}
\begin{proof}
If $\rho(r_{0})=0$ then $e^{\nu (r_{0})}\geq E_{0}$ and by monotonicity, $e^{\nu (r)}\geq E_{0}$ and hence $\rho (r)=0$ for $r>r_{0}$. 
\end{proof}

It is important to control $M$ as the next two lemmas show.

\begin{lemma} If $r\in [0,R_{ex})$ then $M(r)<\frac{1}{8}$.
\end{lemma}
\begin{proof}
Recall the definition of $M$ which reads
\begin{equation*}
M:=\frac{1}{8}(1-e^{2(\nu +\psi -\gamma )}).
\end{equation*}
On the domain of existence $\nu $, $\psi $ and $\gamma $ have finite values, and hence $M<\frac{1}{8}$. 
\end{proof}

We conclude that if $M(\hat{r})=\frac{1}{8}$ then the solution breaks down at $r=\hat{r}$. 
The next corollary follows immediately.

\begin{corollary}
Let $f(r,v):=\phi (E)L^{l}$, $l>-1$ and $k+\frac{l+1}{2}>0$. If $\rho(R)=0$ for some $R$, then a unique vacuum solution, can be joined to the matter cylinder. 
\end{corollary}
\begin{proof}
We have a unique solution on $[0,R]$, since $R<R_{ex}$. Since equations $\left (\ref{s1}\right )$-$\left (\ref{s3}\right )$ are valid in the vacuum case too, and since all functions and their derivatives have finite limits as $r\rightarrow R^{-}$, we can take these values as boundary values for a unique Levi-Civita solution.
\end{proof} If $R_{ex}=+\infty $ the solution is global (in $r$), which in fact is true as the next theorem shows.

\begin{theorem} The system $\left (\ref{s1}\right)$-$\left (\ref{B}\right )$, with $l>-1$, $k+\frac{l+1}{2}>0$ and $\phi $ as in Lemma 1, has a unique global (in $r$) solution, i.e. $R_{ex}=+\infty $.
\end{theorem}
\begin{proof}
Assume that $R_{ex}<+\infty $. First we show that $R_{ex}=R$. Assume that $R<R_{ex}$, then we can glue a unique vacuum solution to the matter solution, but then the solution can not blow up at finite $R_{ex}$ as assumed. Hence, $R=R_{ex}$. 

Now we want to show that $M\in C^{1}([0,R])$. First, $e^{\nu}\leq E_{0}$ since $\nu $ is monotonically increasing. By Lemma 3, $\psi$ and $\nu$ are monotonically decreasing respectively increasing, so by equation $\left (\ref{E2}\right )$ and the boundary conditions $\left (\ref{BC}\right )$ we see that $|\psi |\leq\nu$. Hence, $\psi $ and $\nu $ have finite limits as $r\rightarrow R^{-}$, and $\frac{dM}{dr}=2\pi re^{2\nu}(\frac{4\pi r^{l}e^{-(l+4)\nu -l\psi}}{l+1}[H_{\frac{l+3}{2}}(e^{\nu})+e^{2\nu}H_{\frac{l+1}{2}}(e^{\nu})]-4\pi A_{1}r^{l}e^{-(l+4)\nu-l\psi}H_{\frac{l+3}{2}}(e^{\nu})$ has a finite limit as $r\rightarrow R^{-}$, since $H_{m}\in C^{1}$, $m=\frac{l+1}{2},\frac{l+3}{2}$. Hence, $M\in C^{1}([0,R])$, and we can expand $1-8M(r)=(R-r)S(r)$, with $S\in C([0,R])$. 

Next we prove that $M(r)\leq C<\frac{1}{8}$, $r\in [\frac{R}{2},R)$. By equations $\left (\ref{s2}\right )$ and $\left (\ref{s3}\right )$ we observe that $|\psi '|$ and $\nu '$ are bounded if $M\leq C<1/8$. Hence by equation $\left (\ref{E1}\right )$ $\gamma '$ is also bounded and the solution can be extended contradicting the assumption that $R_{ex}$ is the maximal interval of existence. Recall by Lemma 3, Lemma 4, and $M(0)=0$, that $M(r)\neq 0$, $r\in [\frac{R}{2},R)$, since otherwise there is no matter in spacetime and it becomes a Levi-Civita spacetime with regular axis, i.e. Minkowski spacetime. Now define
\begin{equation}
Q:=\frac{4\pi }{r\sqrt{1-8M}}\int_{0}^{r}\frac{\tilde{r}e^{2\nu}(P_{3}-P_{1})}{\sqrt{1-8M}}d\tilde{r}=\frac{4\pi }{r\sqrt{1-8M}}\int_{0}^{r}\frac{4\pi l\tilde{r}^{l+1}e^{-(l+2)\nu -l\psi}H_{\frac{l+3}{2}}(e^{\nu })}{(l+1)(l+3)\sqrt{1-8M}}d\tilde{r},
\end{equation}
then equations $\left( \ref{s2}\right )$ and $\left (\ref{s3}\right )$ can be written as 
\begin{equation}
\frac{d\psi }{dr}=\frac{1}{2r}[1-\frac{1}{\sqrt{1-8M}}]+Q, \label{psiny}
\end{equation}
\begin{equation}
\frac{d\nu}{dr}=\frac{2}{r(1-8M)}(M+4\pi r^{2}e^{2\nu }P_{1})+Q[rQ-\frac{1}{\sqrt{1-8M}}]. \label{nyny}
\end{equation}

We have to consider two cases. 

First let $-1<l\leq 0$. Then $Q\leq 0$ and hence $Q[rQ-\frac{1}{\sqrt{1-8M}}]\geq 0$. With $r\in [\frac{R}{2},R)$
\begin{equation*}
\frac{d\nu}{dr}=\frac{2}{r(1-8M)}(M+4\pi r^{2}e^{2\nu }P_{1})+Q[rQ-\frac{1}{\sqrt{1-8M}}]\geq\frac{2M(\frac{R}{2})}{1-8M}=\frac{MR}{(R-r)S(r)},
\end{equation*}
and integration yields that $\underset{r\rightarrow R^{-}}{\lim}\nu (r)=+\infty $ which contradicts the fact that $\nu (r)$ is uniformly bounded when $r\in [0,R)$. Hence, $\underset{r\rightarrow R^{-}}{\lim}M(r)<\frac{1}{8}$, and the solution can be extended, contradicting the fact that $R_{ex}$ is the maximal interval of existence. Thus the solution is global (in $r$) when $-1<l\leq 0$.

Now consider the case $l>0$ and define 
\begin{equation*}
\tilde{Q}:=\int_{0}^{r}\frac{\tilde{r}e^{2\nu }(P_{3}-P_{1})}{\sqrt{1-8M}}d\tilde{r}.
\end{equation*}
 By equation $\left (\ref{E2}\right )$ and the boundary conditions $\left (\ref{BC}\right )$
\begin{equation*}
\int_{0}^{r}\frac{\tilde{r}e^{2\nu }P_{1}}{{1-8M}}d\tilde{r}<\nu (r)+\psi (r)<\infty,\text{ }r\in[0,R).
\end{equation*}
Since $P_{3}-P_{2}=lP_{1}$ we have 
\begin{equation*}
\tilde{Q}:=\int_{0}^{r}\frac{\tilde{r}e^{2\nu }(P_{3}-P_{1})}{\sqrt{1-8M}}d\tilde{r}=l\int_{0}^{r}\frac{\tilde{r}e^{2\nu }P_{1}}{\sqrt{1-8M}}d\tilde{r}<\infty,\text{ }r\in[0,R).
\end{equation*}
We observe that $\tilde{Q}(0)=0$ and $\tilde{Q}'(r)>0$, $r\in(0,R)$.

Now by equations $\left (\ref{E2}\right )$ and $\left (\ref{psiny}\right )$
\begin{equation*}
\frac{d\nu }{dr}\geq  r\psi '^{2}-\psi '=\frac{1}{4r}[1-\frac{1}{\sqrt{1-8M}}]^{2}+rQ^{2}+Q[1-\frac{1}{\sqrt{1-8M}}]-\frac{1}{2r}[1-\frac{1}{\sqrt{1-8M}}]-Q
\end{equation*}
\begin{equation*}
=\frac{1}{r(1-8M)}(\frac{1}{4}-4\pi\tilde{Q}+16\pi ^{2}\tilde{Q}^{2})-\frac{1}{4r}.
\end{equation*}

Thus if $\tilde{Q}(r)\leq C<\frac{1}{8\pi }$, $r\in[0,R)$ we have that $\frac{1}{4}+16\pi ^{2}\tilde{Q}^{2}-4\pi\tilde{Q}\geq C>0$. Hence
\begin{equation}
\frac{d\nu }{dr}\geq\frac{C}{r(R-r)S(r)}-\frac{1}{4r},\label{nelly}
\end{equation}
and integration of this inequality from $\frac{R}{2}$ to $r$ and letting $r$ tend to $R^{-}$ will lead to a contradiction.

To obtain this, observe that the integrand in equation $\left (\ref{s2}\right )$, has the same sign so $\psi '<0$, $r\in [0,R)$, and equation $\left (\ref{psiny}\right )$ can be written as
\begin{equation}
\psi '=\frac{1}{2r}+\frac{1}{2r\sqrt{1-8M}}(-1+8\pi\tilde{Q}),\text{ }r\in (0,R). \label{pq}
\end{equation}

Since $\tilde{Q}$ is increasing, $\underset{r\rightarrow R^{-}}{\lim}\tilde{Q}(r)=C$. Now there are three possibilities. First of all assume $C<\frac{1}{8\pi}$, then we are done since this is actually what we want to prove. 

Secondly assume that $C>\frac{1}{8\pi}$. Since $\tilde{Q}$ is continuous, this means that $\tilde{Q}(r)=\frac{1}{8\pi}$, for some $\hat{r}<R$. However, then by equation $\left (\ref{pq}\right )$ $\psi '(\hat{r})>0$, which is a contradiction. 

Hence, we only have to exclude the case $C=\underset{r\rightarrow R^{-}}{\lim}\tilde{Q}(r)=\frac{1}{8\pi }$. Assume by contradiction that $\underset{r\rightarrow R^{-}}{\lim }\tilde{Q}(r)=\frac{1}{8\pi }$, and consider
\begin{equation}
\frac{-1+8\pi\tilde{Q}}{\sqrt{1-8M}}. \label{eps}
\end{equation}
If $\underset{r\rightarrow R^{-}}{\lim}\frac{-1+8\pi\tilde{Q}}{\sqrt{1-8M}}>-1$, then $\psi ' (\hat{r})>0$ for some $\hat{r}\in [0,R)$ by continuity, which is a contradiction. Now by l'Hospital's rule 
\begin{equation*}
\underset{r\rightarrow R^{-}}{\lim}\frac{-1+8\pi\tilde{Q}}{\sqrt{1-8M}}=\underset{r\rightarrow R^{-}}{\lim}\frac{8\pi\tilde{Q}'}{\frac{-4M'}{\sqrt{1-8M}}}
\end{equation*}
\begin{equation*}
=\underset{r\rightarrow R^{-}}{\lim}\frac{\frac{8\pi re^{2\nu(P_{3}-P_{1})}}{\sqrt{1-8M}}}{\frac{-8\pi re^{2\nu }(\rho -P_{1})}{\sqrt{1-8M}}}=\underset{r\rightarrow R^{-}}{\lim }\frac{P_{1}-P_{3}}{\rho -P_{1}}
\end{equation*}
\begin{equation*}
=\underset{r\rightarrow R^{-}}{\lim}\frac{-lH_{\frac{l+3}{2}}(e^{\nu })}{(l+2)H_{\frac{l+3}{2}}(e^{\nu })+(l+3)e^{2\nu }H_{\frac{l+1}{2}}(e^{\nu })}\geq\frac{-l}{(l+2)}>-1.
\end{equation*}
 
Hence we obtained a contradiction, which implies that
\begin{equation*}
\tilde{Q}(r)\leq C<\frac{1}{8\pi },\text{ }r\in [\frac{R}{2},R).
\end{equation*}

Now integrate the inequality $\left (\ref{nelly}\right )$ from $\frac{R }{2}$ to $r$, and let $r$ tend to $R$ we see that $\nu $ diverges to $+\infty$, as $r\rightarrow R^{-}$, which again is a contradiction. Hence $M(r)\leq C<\frac{1}{8},\text{ } r\in[0,R)$ and again the solution is global (in $r$).

\end{proof}

The following corollary is an immediate consequence of Theorem 2. 

\begin{corollary} Let $f(r,v):=\phi (E)L^{l}$, $l>-1$ and $k+\frac{l+1}{2}>0$. Then there exists a unique global (in $r$) $C^{2}$-solution to the system $\left (\ref{E1}\right )-\left (\ref{E4}\right )$ and $\left (\ref{vl}\right )$ subject to the boundary conditions $\left (\ref{BC}\right )$.
\end{corollary}

With the class of distribution functions considered above we could have two possibilities. Namely, either the matter fills all the space or there is a finite radius of extension. In fact as the next theorem shows, the first possibility above is excluded for the class of distribution functions considered in this paper.

\begin{theorem}
Let $f(r,v):=\phi (E)L^{l}$, $l>-1$ and $k+\frac{l+1}{2}>0$. Then there is a finite radius $R$ such that $\rho(r)=0$, $r\geq R$.
\end{theorem}
\begin{proof}
Assume by contradiction that $R=+\infty $. Since we have matter (otherwise $R=0$), and the matter terms are continuous, we can assume that there exists $r_{1}\in (0,+\infty )$ such that $\rho (r_{1})\neq 0$. By continuity of the matter terms $\rho (r)\neq 0$, $r\in (r_{1}-\epsilon ,r_{1}+\epsilon )$ for some $\epsilon >0$. By equation $\left (\ref{s1}\right )$ and Lemma 4 
\begin{equation}
M(r)>\tilde{\epsilon},\label{meps}
\end{equation} 
for some $\tilde{\epsilon}>0$ when $r\geq r_{1}$. We also have 
\begin{equation}
1-\sqrt{1-8M}\geq C>0, r\in [r_{1},+\infty ).\label{meps2}
\end{equation} 
Define
\begin{equation}
\eta :=(\log{E_{0}}-\nu)_{+},
\end{equation}

and
\begin{equation}
\xi :=\frac{M}{\eta}.
\end{equation}
Thus $\eta $ is a well defined and differentiable function since $\log E_{0}:=\nu _{R}$, see equation $\left (\ref{rho}\right )$. Furthermore,
\begin{equation}
\eta '=-\nu ',
\end{equation}
where $'$ denotes differentiation with respect to $r$.
Since $\eta (0)=\nu _{R}>0$, $\xi $ is also a well defined and differentiable function when $r\in [0,R)$.
Now
\begin{equation}
\xi '=\frac{M'}{\eta}-\frac{M\eta '}{(\eta)^{2}}=\frac{M'}{\eta}+\frac{M\nu '}{(\eta)^{2}}\geq \frac{M\nu '}{(\eta)^{2}}=\frac{\nu '}{M}\xi ^{2}.
\end{equation}
By equations $\left (\ref{s1}\right )$ and $\left (\ref{s3}\right )$ we have
\begin{equation}
 \frac{d\nu}{dr}=\frac{4\pi}{r\sqrt{1-8M}}\int_{0}^{r}\tilde{r}e^{2\nu}\frac{1}{\sqrt{1-8M}}(P_{3}+2P_{1}+\rho )d\tilde{r}
\end{equation}
\begin{equation}
=\frac{1}{r\sqrt{1-8M}}[\frac{1}{2}(1-\sqrt{1-8M})+4\pi\int_{0}^{r}\tilde{r}e^{2\nu}\frac{1}{\sqrt{1-8M}}(P_{3}+3P_{1})d\tilde{r}]
\end{equation}
\begin{equation}
\geq\frac{1-\sqrt{1-8M}}{2r\sqrt{1-8M}}.
\end{equation}
By equation $\left (\ref{meps2}\right )$, $1-\sqrt{1-8M(r)}\geq C>0$, $r\in[r_{1},R)$, and thus
\begin{equation}
\xi '\geq\frac{C}{r}\xi ^{2}.
\end{equation}
 
Integrate this inequality from $r_{1}$ to $r$. Then we immediately obtain that there exists $\hat{R}<+\infty$, such that $\xi\rightarrow\infty$ as $r\rightarrow \hat{R}^{-}$. But this implies that $\eta$ tends to zero since $M(r)$ is bounded and increasing. Hence, there exists an $\hat{R}<\infty$ such that $\rho(r)=0$, $r\geq \hat{R}$ which contradicts the assumption above.
\end{proof}

\subsection{Perfect fluid}

In this section we will consider the Einstein equations coupled to a perfect fluid. By definition, a barotropic fluid is one whose energy density and pressure are related by an equation of state that does not contain the temperature. We will assume the fluid to have a barotropic equation of state

\begin{equation}
\rho =\rho (P),
\end{equation}

where $\rho $ and $P$ are the energy density respectively the isotropic pressure of the fluid. We will also assume that the equation of state $\rho =\rho (P)$, is $C^{1}$ when $P>0$, and $\rho (P)\rightarrow \rho_{R}<+\infty,\text{ }P\rightarrow 0$, where $\rho _{R}\geq 0$ is a constant. 

Many of the results in this section is already known, cf. \cite{BL}. However, for completeness we sketch the proof of the results. The proofs essentially follows the proofs in section 3.1 for the corresponding results. The important new result of this section is that we are able to treat a more general class of equations of state than what is done in \cite{BL}. Hence Theorem 5 is the main result of this section. The following local existence result follows from Lemma 7 below.

\begin{proposition}
Let $\rho (P)$, be a barotropic equation of state, $\rho\in C^{1}((0,P_{c}])$, where $P _{c}$ is the central pressure and furthermore $\underset{P\rightarrow 0}{\lim }\rho (P)=\rho _{R}$, where $\rho _{R}<+\infty $ is a nonnegative constant. For each value of the central density $\rho _{c}=\rho (P_{c})>0$, there exists $r_{0}>0$, such that a unique $C^{2}$-solution of the field equations $\left (\ref{E1}\right )$-$\left (\ref{E5}\right )$ with boundary conditions $\left (\ref{BC}\right )$, exists for $r\in [0,r_{0}]$.
\end{proposition}

In the perfect fluid case with $M:=\frac{1}{8}(1-e^{2(\nu +\psi -\gamma )})$ as in section 3.1, the equations corresponding to equations $\left (\ref{s1}\right )$-$\left (\ref{s3}\right )$ can be obtained by observing that the perfect fluid case corresponds to the $l=0$ case for Vlasov matter. Hence, $P_{1}=P_{2}=P_{3}:=P$ in this case, and by using equation $\left (\ref{s1}\right )$ it is possible to integrate equations $\left (\ref{s2}\right )$ and $\left (\ref{s3}\right )$ to obtain equations $\left (\ref{ss2}\right )$ and $\left (\ref{ss3}\right )$. The analogue of equations $\left (\ref{s1}\right )$-$\left (\ref{B}\right )$ with equation $\left (\ref{E5}\right )$ added are 

\begin{equation}
\frac{dM}{dr}=2\pi re^{2\nu }(\rho -P),\label{ss1}
\end{equation}
\begin{equation}
\frac{d\psi }{dr}=\frac{1}{2r}[1-\frac{1}{\sqrt{1-8M}}],\label{ss2}
\end{equation}
\begin{equation}
\frac{d\nu }{dr}=\frac{2}{r(1-8M)}(M+4\pi r^{2}e^{2\nu }P),\label{ss3}
\end{equation}
\begin{equation}
\frac{dP}{dr}=-\frac{d\nu }{dr}(P+\rho ),\label{ss4}
\end{equation}
\begin{equation}
M=\frac{dM}{dr}=\nu=\frac{d\nu}{dr}=\psi=\frac{d\psi}{dr}=0,\text{ at } r=0,P(0)=P_{c}>0. \label{Bf}
\end{equation}

Then we have the following result.

\begin{lemma}
Let $\rho (P)$, be a barotropic equation of state, $\rho\in C^{1}((0,P_{c}])$, where $P _{c}$ is the central pressure and furthermore $\underset{P\rightarrow 0}{\lim }\rho (P)=\rho _{R}$, where $\rho _{R}<+\infty $ is a nonnegative constant. For each value of the central density $\rho _{c}=\rho (P_{c})>0$, there exists $r_{0}>0$, such that a unique $C^{1}$-solution of the equations $\left (\ref{ss1}\right )$-$\left (\ref{ss4}\right )$ with boundary conditions $\left (\ref{Bf}\right )$, exists for $r\in [0, r_{0}]$.
\end{lemma}
\begin{proof}
The proof follows the proof of Lemma 2 with $l=0$. 
\end{proof} 

From equations $\left (\ref{E1}\right )$, and $\left (\ref{ss1}\right )-\left (\ref{ss4}\right )$ we can easily obtain the following lemma.
\begin{lemma}
Suppose we have a solution with a regular axis. Then $M$, $\nu$ and $\gamma$ are strictly increasing and $\psi$ and $P$ are strictly decreasing as long as $P>0$.
\end{lemma}
\begin{proof}
Using equations $\left (\ref{E1}\right )$, $\left (\ref{ss1}\right )$-$\left (\ref{ss4}\right )$ we can immediately read off the claimed properties of the different functions.
\end{proof}

The following is an easy consequence of equation $\left (\ref{ss4}\right )$. We recall the definitions 
\begin{equation}
R_{ex}:=\text{ the maximal interval of existence, }
\end{equation}
 and 
\begin{equation}
R:=\sup\{r\leq R_{ex}:\rho (\tilde{r})>0,\text{ for }\tilde{r}\in[0,r)\}.
\end{equation}

The following lemma holds also in the perfect fluid case.

\begin{lemma} Assume that there exists $r_{0}\in (0,R_{ex})$ such that $\rho (r_{0})=0$, then $\rho (r)=0$ for $r\in[r_{0},R_{ex})$.
\end{lemma}
\begin{proof}
Let $r_{0}$ be as above. Then $\rho (r_{0})=0$ and since $P\leq \rho $, $P(r_{0})=0$. By equation $\left (\ref{ss4}\right )$, $P$ is a decreasing function, thus $P(r)=0$, $r\in [r_{0},R_{ex})$. But $\rho $ considered as a function of $P$ implies that $\rho (r)=0$, $r\in [r_{0},R_{ex})$.
\end{proof}

Lemma 5 and Lemma 6 holds without any changes in the proofs.

\begin{lemma} If $r\in [0,R_{ex})$, then $M(r)<\frac{1}{8}$.
\end{lemma}

 The analogue of Theorem 2 holds for perfect fluid matter since we have equation $\left (\ref{ss3}\right )$ which reads
\begin{equation*}
\frac{d\nu }{dr}=\frac{2}{r(1-8M)}(M+4\pi r^{2}e^{2\nu }P).
\end{equation*}
Then it is easy to apply the same strategy as in the proof of Theorem 2. We have the following corollary which says that we can extend the spacetime by adding a unique vacuum solution to it. 
\begin{corollary}
If $P(R)=0$, for some $R<\infty $, then a unique vacuum solution can be joined to the fluid cylinder.
\end{corollary}
\begin{proof}
See the sketch of proof of Corollary 2.
\end{proof}

\begin{theorem}
Let $\rho (P)$, be a barotropic equation of state, $\rho\in C^{1}((0,P_{c}))$, where $P _{c}$ is the central pressure and furthermore $\underset{P\rightarrow 0}{\lim }\rho (P)=\rho _{R}$, where $\rho _{R}<+\infty $ is a nonnegative constant. Then for each value of the central density $\rho _{c}=\rho (P_{c})>0$, there exists a unique global (in $r$), $C^{1}$-solution of the equations $\left (\ref{ss1}\right )$-$\left (\ref{ss4}\right )$ with boundary conditions $\left (\ref{Bf}\right )$.
\end{theorem}
 The following corollary is an immediate consequence of Theorem 4.

\begin{corollary} Let $\rho (P)$ be a barotropic equation of state, then there exists a unique global (in $r$) $C^{2}$-solution to the system $\left (\ref{E1}\right )-\left (\ref{E5}\right )$ subject to the boundary conditions $\left (\ref{BC}\right )$ and $0<P_{c}<+\infty $.
\end{corollary} 
Now we have two possibilities. Either the fluid fills all the space or there is a finite radius where the fluid vanishes. 

In \cite{BL} they proved that for perfect fluids with positive boundary densities  $\rho (0)>0$, and regular axis, there always exists $R<\infty$, such that $P(R)=0$. This fact can be generalized to all equations of state which satisfy 
\begin{equation*}
\int_{0}^{P_{c}}\frac{dp}{\rho (p)+p}<\infty.
\end{equation*}
Hence, we have the following theorem

\begin{theorem}
For a solution with regular axis and an equation of state that satisfies $\rho _{0}=\rho (P_{c})>0$ and
\begin{equation}
\int_{0}^{P_{c}}\frac{dp}{\rho (p)+p}<\infty,\label{aseta}
\end{equation}
there is a finite radius $R$ such that $P(R)=0$.
\end{theorem}
\begin{proof}
The proof follows the proof of Theorem 2, but we outline it for completeness. Define 
\begin{equation}
\eta (P):=\int_{0}^{P}\frac{dp}{\rho (p)+p}, \label{eta}
\end{equation}
and
\begin{equation}
\xi :=\frac{M}{\eta}.
\end{equation}
Observe that $\xi $ is well defined, since it is matter at the axis of symmetry and the equation of state is continuous. Furthermore $\eta $ is well-defined and $\eta (P_{c})$ is finite and nonzero. We also have that $\eta $ is continuous, and $M$ is finite. By $\left (\ref{ss4}\right )$
\begin{equation}
\eta '=-\nu ',
\end{equation}
where $'$ denotes differentiation with respect to $r$.
Now $M(r)>\tilde{\epsilon }$, $r\in [r_{1},+\infty )$, $\tilde{\epsilon } >0$ (see the proof of Theorem 3, section 3.1) so by equation $\left (\ref{ss3}\right )$
\begin{equation}
\xi'=\frac{M'}{\eta}-\frac{M\eta '}{(\eta)^{2}}=\frac{M'}{\eta}+\frac{M\nu '}{(\eta)^{2}}\geq \frac{M\nu '}{(\eta)^{2}}=\frac{\nu '}{M}\xi^{2}\geq\frac{C}{r}\xi^{2}.
\end{equation}
From this inequality we immediately obtain that there exists $R<\infty$, such that $\xi\rightarrow\infty $, as $r\rightarrow R^{-}$. But this implies that $\eta$ tends to zero since $M<\frac{1}{8}$. Hence, there exists an $R<\infty$, such that $P(R)=0$.
\end{proof}

\begin{remark}
Observe that $\eta$ is not bounded for linear equations of state of the type $P=(\mu -1)\rho,\text{ } 1\leq\mu\leq 2$ which is in complete agreement with the explicit solutions obtained in \cite{K}. In \cite{SST} however they consider polytropic equations of state $\rho =(\frac{P}{K})^{\frac{1}{\mu}}+nP,\text{ } n=\frac{1}{1-\mu}$ where $K$ and $\mu$ are constants, and these models have finite radius as can be seen by evaluation of the integral $\left (\ref{eta}\right )$ in the definition of $\eta$. Also observe that for equations of state with positive boundary density, i.e. $\rho (0)>0$, the integral $\left (\ref{eta}\right )$ is finite, compare with the results in \cite{BL}.
\end{remark} 

\begin{remark}
The proof of finiteness of the radius is much easier than in \cite{M} because we only need to consider one equation and all terms turn out to have the right sign. However, if one tries to apply exactly the same strategy in the spherically symmetric case it fails due to an $r^{-2}$-term that appears in the differentiation of $\eta $. 
\end{remark}

\subsection{The Vlasov-Poisson system}
 Consider now the system $\left (\ref{vp1}\right )-\left (\ref{vp2}\right )$, which with the ansatz $f(r,v)=\phi (E)L^{l},\text{ } l>-1$, and $\phi $ satisfies the assumptions in Lemma 1, is turned into a nonlinear equation for $U$

\begin{equation}
(rU')'=\frac{2^{\frac{l+9}{2}}\pi ^{2}}{l+1} r^{l+1}g_{\frac{l+1}{2}}(U), \label{VP}
\end{equation}

where

\begin{equation*}
g_{m}(U):=\int_{U}^{E_{0}}\phi (E)(E-U)^{m}.
\end{equation*}

Let $U$ satisfy the boundary conditions $U(0)=a,\text{ }U'(0)=0$, where $a$ is a constant. Then we have the following local existence result, cf. \cite{BB} where they prove a more general existence theorem.

\begin{proposition}
Let $f(r,v)=\phi (E)L^{l},\text{ }l>-1$, and $\phi $ as in Lemma 1 with $k+\frac{l+1}{2}>0$ such that $g_{\frac{l+1}{2}}\in C^{1}((-\infty,+\infty ))$. Then there exists $\delta >0$, such that equation $\left (\ref{VP}\right )$, subject to the boundary conditions $U(0)=a,\text{ }U'(0)=0$, has a unique $C^{1}$-solution when $r\in [0,\delta]$.
\end{proposition}
\begin{proof}
The proof is omitted since it is similar to the proof of Lemma 2.
\end{proof}

The solution is global as the next theorem shows, cf. \cite{BB} for a more general theorem.

\begin{proposition}
Let $f(r,v)=\phi (E)L^{l},\text{ }l>-1$, and $\phi $ as in Lemma 1 with $k+\frac{l+1}{2}>0$ such that $g_{\frac{l+1}{2}}\in C^{1}((-\infty,\infty ))$. Then there exists a unique $C^{1}$-solution of equation $\left (\ref{VP}\right )$, subject to the boundary conditions $U(0)=a,\text{ }U'(0)=0$, on $[0,\infty )$.
\end{proposition}
\begin{proof}
Let $R_{ex}$ and $R$ be defined as in the relativistic case. Assume that $R_{ex}<+\infty $. Clearly $U$ is increasing. Either $\underset{\lim}{r\rightarrow R_{ex}}U(r)\leq E_{0}$ on its maximal interval of existence, which implies that
\begin{equation}
g_{\frac{l+1}{2}}(U)\leq C,
\end{equation}
 by Lemma 1, and by equation $\left (\ref{VP}\right )$
\begin{equation}
U'(r)\leq Cr^{l+1},
\end{equation}
which is uniformly bounded on $[0,R_{ex})$, since $l>-1$. Thus the solution can be extended which contradicts the maximality of the solution, and hence the solution is global (in $r$).

If on the other hand $\underset{\lim}{r\rightarrow R_{ex}}U(r)>E_{0},\text{ }r\in [0,R_{ex})$, then for all $r\geq R$, $\rho (r)=0$, and again $U$ exists globally since we can add a vacuum solution to the matter solution at $r=R$. 
\end{proof} 

In fact with $f$ as above the matter has finite extension which we show below.

\begin{theorem}
Let $f(r,v):=\phi (E)L^{l}$, $l>-1$, $k+\frac{l+1}{2}>0$ and $\phi $ be as in Lemma 1, then there exists an $R$, such that $\rho (r)=0$, $r\geq R$.
\end{theorem}

\begin{proof}
Define 
\begin{equation*}
m:=2\pi\int_{0}^{r}\tilde{r}\rho (\tilde{r}) d\tilde{r},
\end{equation*}
 and 
\begin{equation*}
\eta:=(E_{0}-U)_{+},
\end{equation*}
 then $m$ and $\eta $ are well-defined and satisfy the following equations

\begin{equation*}
m'=2\pi r\rho,
\end{equation*}
and 
\begin{equation*}
\eta '=-\frac{2m}{r},
\end{equation*}
where $'$ denotes differentiation with respect to $r$.
 Now define 
\begin{equation*}
\xi:=\frac{m}{\eta},
\end{equation*}
then $\xi$ satisfies
\begin{equation*}
\xi '=\frac{m'}{\eta}-\frac{m\eta '}{(\eta)^{2}}\geq \frac{2}{r}\xi ^{2}.
\end{equation*}
As in the relativistic case, there exist an $R$ such that $\xi(r)\rightarrow\infty$, as $r\rightarrow R^{-}$. Hence there exists $R$ such that $\rho (r)=0$, $r\geq R$.
\end{proof} 
 
\textbf{Acknowledgment:} I want to thank Håkan Andreasson for useful discussions and comments.

\end{document}